\documentclass[11pt,letterpaper]{article}

\usepackage[textwidth=6in,top=1in,bottom=1in,includeheadfoot,headheight=14pt,headsep=12pt,footskip=22pt]{geometry}
\usepackage{lmodern}
\usepackage{fontspec}
\usepackage{microtype}
\usepackage{graphicx}
\usepackage{booktabs}
\usepackage{amsmath}
\usepackage{tikz}
\usetikzlibrary{arrows.meta,positioning,backgrounds}
\usepackage[font=small,labelfont=bf]{caption}
\usepackage[super,sort&compress]{natbib}
\setcitestyle{super,comma}
\usepackage[section]{placeins}
\usepackage{flafter}
\usepackage{xurl}
\usepackage[hidelinks]{hyperref}
\usepackage{textcomp}

\graphicspath{{figures/}}
\hypersetup{
  pdftitle={A Heckler in the Hidden State: Correctness Signals in Diffusion Language Models},
  pdfauthor={Angad Miglani; Samrath Singh Chadha; Kevin Li; Manas Venkata Sai Ravulapalli},
  pdfsubject={Correctness probes and activation steering in diffusion language models},
  pdfkeywords={diffusion language models, activation steering, linear probes, code generation}
}
\definecolor{accent}{HTML}{0F4D92}
\definecolor{emph}{HTML}{B64342}
\definecolor{ink}{HTML}{1A1A1A}
\definecolor{soft}{HTML}{5A6472}
\tikzset{
  dbox/.style={rounded corners=3pt,draw=accent!65,fill=accent!5,line width=0.8pt,
    align=center,inner sep=7pt,minimum height=1.2cm,text width=3.5cm,font=\small},
  dhl/.style={dbox,draw=emph!65,fill=emph!6},
  darr/.style={-{Stealth[length=2.4mm]},line width=0.9pt,draw=soft}
}

\title{\vspace{-2em}\LARGE\bfseries A Heckler in the Hidden State:\\Correctness Signals in Diffusion Language Models}
\author{%
  {\normalfont\fontsize{10}{12}\selectfont
    \begin{tabular}{@{}c@{\hspace{0.65em}}c@{\hspace{0.65em}}c@{\hspace{0.65em}}c@{}}
      Angad Miglani & Samrath Singh Chadha & Kevin Li & Manas Venkata Sai Ravulapalli\\[3pt]
      {\small\href{mailto:angad.miglani@duke.edu}{\nolinkurl{angad.miglani@duke.edu}}} &
      {\small\href{mailto:samrath@perseus.so}{\nolinkurl{samrath@perseus.so}}} &
      {\small\href{mailto:kevin.li3@duke.edu}{\nolinkurl{kevin.li3@duke.edu}}} &
      {\small\href{mailto:manas@perseus.so}{\nolinkurl{manas@perseus.so}}}
    \end{tabular}}\\[12pt]
  {\normalfont\normalsize Efficient Computation Inc.}
}
\date{}

\begin{document}

\maketitle
\thispagestyle{empty}
\vspace{-2em}

% ================= ABSTRACT =================
\begin{abstract}
Diffusion language models generate code by repeatedly updating a partially masked sequence. We ask whether their internal activations encode code correctness and whether that information can improve generation. Across six diffusion models, linear probes distinguish passing from failing attempts, with the strongest reads generally appearing beyond the early layers. Controls using small semantic mutations support a connection to correctness rather than surface style alone. In comparisons with model confidence, the probes offer no consistent advantage. Adding a probe-derived direction to the residual stream does not yield a dependable improvement in the tested steering settings, while the opposite direction degrades performance. The models carry information about correctness, but the tested interventions do not turn that information into better code.
\end{abstract}

\section{Introduction}\label{sec:introduction}

A model can contain information that its output does not make useful. For code generation, the distinction is concrete: an activation may help predict whether a program will pass its tests without providing an intervention that repairs the program. Work on language-model self-evaluation shows that models can sometimes estimate whether their answers are correct, although calibration depends on the task and prompting.\citep{kadavath2022} This motivates a separate question about internal representations: does a readable correctness signal also provide a way to improve generation?

Diffusion language models offer a useful setting for this question. Instead of committing to one token at a time, they revise positions in a partially masked sequence over several denoising steps. A correctness-related representation could therefore appear while the program can still change. The difficulty is that a probe may read the consequences of a successful computation without identifying how to perform that computation. For example, a direction that distinguishes a correct loop condition from an incorrect one may still give no indication of which operator should be changed.

We study three questions in frozen code models. First, can a linear probe distinguish correct and incorrect attempts at the same programming problem? Second, does it add information beyond the model's own confidence? Third, can a direction fitted to correctness labels improve the generated code when added to the model's activations? We compare six diffusion models, run additional checks on DiffuCoder-7B, and extend the comparison to autoregressive code models.

Correctness is readable in the tested diffusion models, but the probes offer no consistent advantage over the model's own confidence. Steering along those directions also fails to produce a dependable gain: pushing toward the passing direction leaves performance unchanged or lower, while pushing in the opposite direction causes larger losses. We examine this asymmetry through confidence controls, activation perturbations, and several steering procedures.

\section{Background and related work}\label{sec:background}

\subsection{Diffusion and language generation}

Sohl-Dickstein and colleagues introduced generative modelling through a forward process that gradually corrupts data and a learned reverse process that reconstructs it.\citep{sohldickstein2015} Ho and colleagues developed the denoising diffusion probabilistic model formulation that became widely used for continuous data.\citep{ho2020} Language requires a treatment of discrete tokens. Austin and colleagues studied structured discrete diffusion processes, including transitions involving an absorbing state.\citep{austin2021} Masked language diffusion uses this idea to generate by repeatedly predicting and updating masked positions.

LLaDA demonstrated large-scale language modelling with a masked diffusion objective.\citep{nie2025} DiffuCoder adapted masked diffusion to code and studied how its generation order and reinforcement-learning procedure affect code synthesis.\citep{gong2025} We study these models with their weights frozen, reading and editing internal states while multiple output positions remain revisable.

\subsection{Reading representations and changing behaviour}

Linear probes test whether a representation makes a target label accessible to a simple classifier. Alain and Bengio used linear classifier probes to study representations across network depth.\citep{alain2016} Here the label is whether a generated program passes its tests, and the probe is a difference-of-means direction. This gives us an explicit score for reading correctness and a direction we can subsequently test by steering.

Activation steering changes model behaviour by adding a direction to its internal activations. Turner and colleagues demonstrated this approach in language models.\citep{turner2023} For masked diffusion models, Shnaidman and colleagues studied steering refusal behaviour,\citep{shnaidman2026} and Zhou and colleagues studied attribute-dependent schedules for interventions during denoising.\citep{zhou2026} We apply this approach to code correctness, testing both the probe direction and a direction optimized for reference-token probability.

The distinction also appears in studies of autoregressive models. Roy and colleagues report that probe-derived directions can detect hallucination-related signals while failing to correct them in their tested settings.\citep{roy2026} Basu and colleagues examine a related gap between internal discrimination and successful intervention in a clinical triage task.\citep{basu2026} Our study examines this question in diffusion code generation, with unit-test labels, confidence controls, perturbations, and direct generation interventions.

\section{Methods}\label{sec:methodology}

\subsection{Tasks, models, and correctness labels}

We evaluate six diffusion code models: DiffuCoder-7B, Dream-Coder-7B, Stable-DiffCoder-8B, UltraLLaDA-8B, LLaDA2.0-mini, and LLaDA-flash. LLaDA2.0-mini has approximately 1.4B active parameters, whereas LLaDA-flash has approximately 102B total parameters; these counts describe different quantities. All model weights remain frozen. The main programming benchmark is MBPP+, with additional mutation controls on HumanEval+, using the EvalPlus evaluation framework.\citep{liu2023}

We label an attempt correct when it passes the available unit tests. Probing uses mixed problems, for which at least one attempt passes and at least one fails. Comparing attempts at the same problem prevents task difficulty alone from determining the ranking.

The family comparison and follow-up experiments use separate sets of generated attempts. The initial DiffuCoder study contains 320 attempts from 64 mixed problems. After mixed-problem filtering, the 32-step confidence comparison contains 95 attempts from 19 problems, and the Qwen2.5-Coder comparison contains 448 attempts from 56 problems. Appendix~\ref{app:methods} gives the generation settings and sample sizes.

\subsection{Probes and confidence controls}

We refer to the response-token-pooled residual-stream activation at a layer as a J-space vector $h$. For a collection of labelled attempts, the probe direction is
\[
v^{*}=\operatorname{mean}(h\mid\mathrm{pass})-\operatorname{mean}(h\mid\mathrm{fail}).
\]
The score of an attempt is its projection onto that direction. The initial family comparison fits and evaluates directions on separate problem sets. In the follow-up experiments, activations are centered within each problem and directions are fitted in five folds grouped by problem. All attempts at a given problem remain in the same fold.

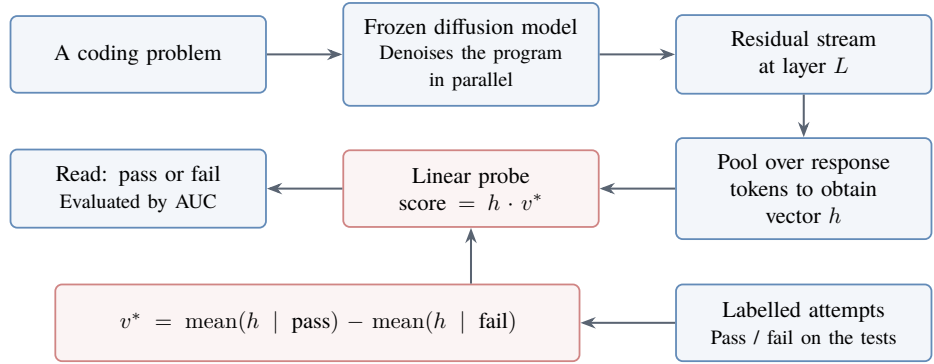
\begin{figure}[htbp]
\centering
\resizebox{0.80\linewidth}{!}{%
\begin{tikzpicture}
  \node[dbox] (p) at (0,0) {A coding problem};
  \node[dbox] (m) at (5.2,0) {Frozen diffusion model\\\footnotesize Denoises the program\\in parallel};
  \node[dbox] (r) at (10.4,0) {Residual stream\\at layer $L$};
  \node[dbox] (pool) at (10.4,-2.1) {Pool over response\\tokens to obtain\\vector $h$};
  \node[dhl] (probe) at (5.2,-2.1) {Linear probe\\$\text{score}=h\cdot v^{*}$};
  \node[dbox] (read) at (0,-2.1) {Read: pass or fail\\\footnotesize Evaluated by AUC};
  \draw[darr] (p) -- (m);
  \draw[darr] (m) -- (r);
  \draw[darr] (r) -- (pool);
  \draw[darr] (pool) -- (probe);
  \draw[darr] (probe) -- (read);
  \node[dbox] (train) at (10.4,-4.2) {Labelled attempts\\\footnotesize Pass / fail on the tests};
  \node[dhl,text width=7.7cm] (vstar) at (2.8,-4.2)
    {$v^{*}=\operatorname{mean}(h\mid\text{pass})-\operatorname{mean}(h\mid\text{fail})$};
  \draw[darr] (train) -- (vstar);
  \draw[darr] (vstar.north -| probe.south) -- (probe.south);
\end{tikzpicture}}
\caption{Read pipeline. Only the pooled residual-stream vector $h$ at one layer is read. The probe direction $v^{*}$ is built once from labelled attempts (bottom) and then
applied to held-out attempts (top). Results are compared with explicit null baselines: shuffled pass/fail labels for probing and random vectors for steering.}
\label{fig:pipeline}
\end{figure}

Within-problem area under the receiver operating characteristic curve (AUC) measures how often a passing attempt receives a higher score than a failing attempt for the same problem. Scores are averaged across mixed problems. An AUC of 0.5 corresponds to chance ranking, and 1.0 to perfect ranking. We also compare probe scores with model-derived confidence measures, including token entropy and completion log probability. Figure~\ref{fig:pipeline} shows the read pipeline.

We report the maximum AUC across the out-of-fold layer scores. Layer selection uses these same scores, without an additional holdout. Appendix~\ref{app:uncertainty} specifies the permutation and resampling procedures and their limits.

\subsection{Perturbations and steering}

We measure probe AUC after random residual displacements and attention-edge ablations. For DiffuCoder, the noise experiments also measure pass@1 on generated code. Attention-edge ablations evaluate the probe only. Steering experiments generate new completions and score them with unit tests.

Steering adds a normalized probe direction to response-token activations, scaled relative to the residual norm. Controls include an untreated baseline and random directions. The follow-up experiments also test interventions restricted to still-masked positions and a learned direction optimized for reference-token log probability. All eleven conditions run as rows of one generation batch on 30 evaluation tasks; exact settings and counts appear in Appendix~\ref{app:steering}.

\section{Results}\label{sec:experiments}

\subsection{Correctness is readable beyond surface features}\label{sec:representations}

\begin{figure}[htbp]
\centering
\includegraphics[width=.74\linewidth]{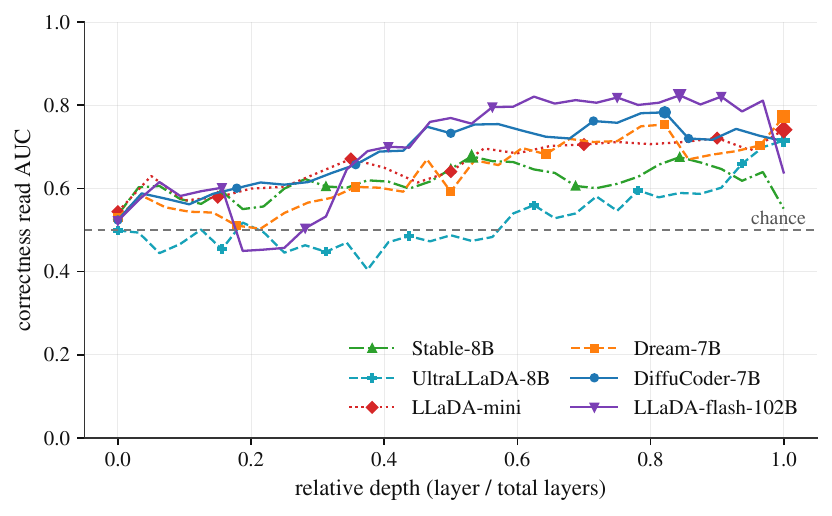}
\caption{Correctness-probe AUC across depth in six frozen diffusion code models. The signal strengthens beyond the input layers and peaks at different depths. These curves are reconstructed from result summaries; Appendix~\ref{app:models} explains their differences from the original peak estimates, and Appendix~\ref{app:final} compares final-block and normalized states.}
\label{fig:depth}
\end{figure}

Every diffusion model in the family comparison has a peak AUC above 0.5. The depth curves peak at approximately 0.68 for Stable-DiffCoder and 0.82 for LLaDA-flash, the endpoints of the range. Correctness becomes more readable beyond the input layers, but the location and shape of the peak differ across models (Fig.~\ref{fig:depth}).

A probe could succeed by recognizing surface differences between passing and failing code. The mutation controls test this explanation with small semantic changes, such as replacing \texttt{<} with \texttt{<=}, while retaining compilable code and matching mutation kind. DiffuCoder distinguishes the resulting pass/fail pairs with AUCs of 0.670 on 392 MBPP+ mutants and 0.692 on 334 HumanEval+ mutants. An earlier test on mutations of the model's own code gives an AUC of 0.774. The read therefore survives when passing and failing programs differ by small edits rather than broad changes in style.

Readability also differs from interpretability through output tokens. Projecting the direction through the output embedding yields an explicit error-related vocabulary for the LLaDA2-MoE models, but little such vocabulary for the dense models. That contrast is shown in Appendix~\ref{app:models}. A direction can discriminate pass/fail labels even when its token-level interpretation is unclear.

\subsection{Confidence provides a competitive readout}

DiffuCoder gives a probe AUC of 0.842, compared with 0.838 for negative token entropy and 0.798 for the masked-forward confidence score. On the same attempts within the Qwen2.5-Coder experiment, probe AUC is 0.677 and log-probability confidence AUC is 0.703 (Table~\ref{tab:confidence}). The probe is slightly ahead of the strongest confidence baseline for DiffuCoder and behind it for Qwen2.5-Coder.

\begin{table}[htbp]
\centering\small
\caption{Probe and confidence AUC on mixed problems. Each model's probe and confidence scores use the same attempts; the two models use different cohorts (Appendix~\ref{app:methods}).}
\label{tab:confidence}
\begin{tabular}{@{}lrrl@{}}
\toprule
Model & Probe AUC & Confidence AUC & Confidence measure\\
\midrule
DiffuCoder-7B & 0.842 & 0.838 & Negative token entropy\\
DiffuCoder-7B & 0.842 & 0.798 & Masked-forward score\\
Qwen2.5-Coder & 0.677 & 0.703 & Completion log probability\\
\bottomrule
\end{tabular}
\end{table}

Model confidence is thus a competitive baseline for the correctness readout. Reading hidden states provides little apparent gain in these comparisons over scores obtained directly from the output distribution. The DiffuCoder comparison contains only 19 mixed problems; uncertainty in that comparison is discussed in Section~\ref{sec:limits}.

\subsection{Noise affects the read more than generation}

\begin{figure}[htbp]
\centering
\includegraphics[width=.60\linewidth]{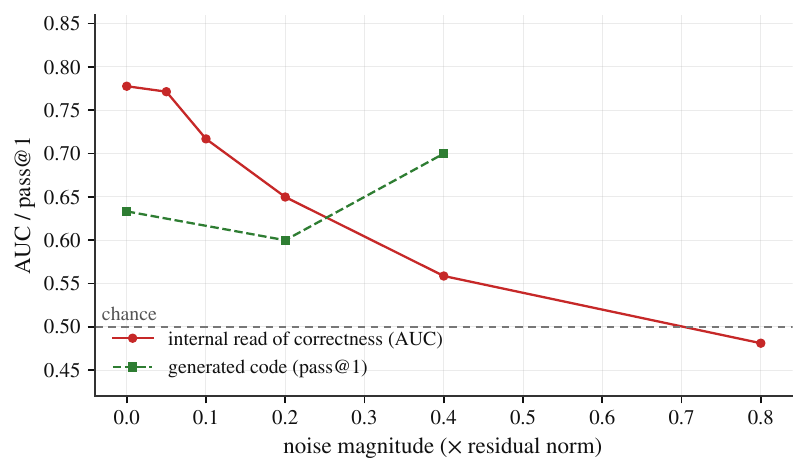}
\caption{DiffuCoder probe AUC and generation pass@1 under noise. The curves cover different ranges of noise magnitude. Over the measured behavioural range, probe AUC falls while pass@1 remains comparatively stable.}
\label{fig:dissociation}
\end{figure}

Random residual noise reduces DiffuCoder's probe AUC, while pass@1 remains comparatively stable over the measured behavioural range (Fig.~\ref{fig:dissociation}). The read is more sensitive to these perturbations than the generated code.

Appendix~\ref{app:perturbations} extends the noise comparison to three models with five seeds each and shows the attention-ablation results. Both perturbations reduce probe performance, with different rates of decline across models.

\subsection{Tested steering procedures do not provide a dependable gain}

\begin{figure}[htbp]
\centering
\includegraphics[width=.60\linewidth]{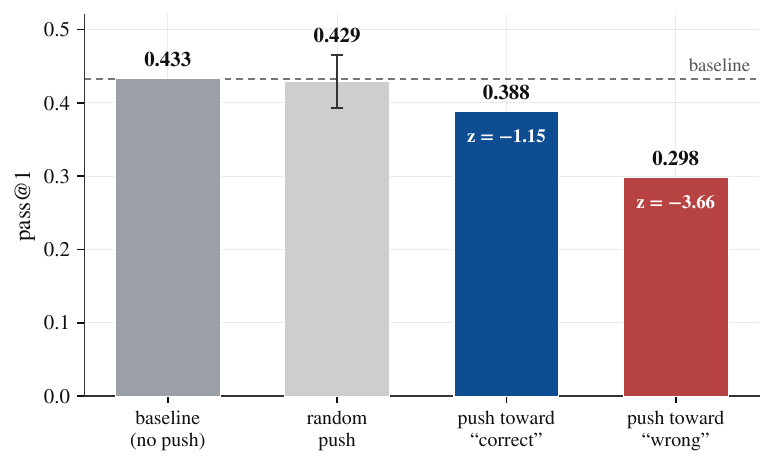}
\caption{DiffuCoder steering results. The positive probe direction lowers pass@1, and the negative direction causes a larger loss. The random-direction error bar is $\pm1$ standard deviation across ten directions; $z$ measures displacement from their mean in standard-deviation units.}
\label{fig:steering}
\end{figure}

The initial steering experiment compares the probe direction with ten random directions of matched norm. Pass@1 is 0.433 without intervention, 0.429 averaged over random directions, 0.388 when pushed toward the passing direction, and 0.298 when pushed toward the failing direction (Fig.~\ref{fig:steering}). Both probe interventions lower performance, with the larger loss in the negative direction.

The follow-up experiments broaden the interventions. The baseline solves 15 of 30 tasks, a global positive probe intervention solves 14, and its negative counterpart solves 5. Restricting the positive intervention to masked positions matches the baseline at 15 tasks. Neither positive probe intervention improves on the untreated model.

Optimizing a separate direction for reference-token log probability also lowers performance. Its global positive arm passes 5 tasks, compared with 15 for the baseline, and compiles on 9 rather than 26 tasks. The learned direction has cosine similarity 0.108 with the probe direction: optimization moves it nearly orthogonal to the original read, yet still fails to improve unit-test outcomes. Appendix~\ref{app:steering} gives the full eleven-condition comparison.

\section{Discussion and limitations}\label{sec:discussion}

\subsection{A useful readout need not be a useful intervention}

The steering results suggest that a general correctness verdict may be too coarse to repair code. A failing program may require a different boundary condition, a missing update, or a change in algorithm. These errors share a pass/fail label but require different edits. An intervention tied to the location and type of error may therefore be more useful than a single direction averaged across many failures.

Another possibility is that later computation modifies or compensates for the activation edit. Both explanations fit the observed asymmetry: a perturbation can damage working code without supplying what a broken program needs. Successful refusal and attribute steering in diffusion models provides a reason to investigate more specific targets and denoising schedules.\citep{shnaidman2026,zhou2026}

\subsection{Depth and architecture require matched comparisons}\label{sec:comparison}

Is a readable correctness signal special to diffusion models? The autoregressive comparisons suggest otherwise: Qwen3-Coder supports a clear read, while GLM-4.5-Air is much closer to its shuffled-label baseline. In the Qwen2.5-Coder comparison, the probe reaches an AUC of 0.677 and the model's own log-probability confidence reaches 0.703. Even with three autoregressive models, architecture cannot be separated from training. Appendix~\ref{app:architecture} develops this comparison and the different opportunities for using a signal during generation.

The final-layer measurements resist a simple scaling explanation. The LLaDA-flash curve prompted a compression hypothesis, but DiffuCoder's read remains strong at the final block. Stable-DiffCoder also shows a final-layer decline, so the pattern is not confined to the 102B model. Appendix~\ref{app:final} compares the raw and normalized states and the different measurement protocols.

\subsection{Limitations}\label{sec:limits}

The experiments use different groups of tasks and generation settings. Probe evaluation is restricted to mixed problems, and the 16-step and 32-step DiffuCoder experiments contain different cohorts. Their AUC difference therefore cannot be attributed to denoising steps alone. Exact reruns also require capture helpers and activation arrays that are not included in the supplementary package, as well as model and dataset revisions that were not consistently recorded.

The follow-up permutation analysis holds fitted probe scores fixed while shuffling labels. Its bootstrap resamples problem identifiers but merges repeated draws, losing the multiplicity required for a cluster bootstrap. Layer selection also lacks a separate holdout, and AUC ranks do not explicitly average ties. We therefore compare point estimates rather than make significance or equivalence claims from those uncertainty estimates. Appendix~\ref{app:uncertainty} gives the procedures; recalibration requires refitting under permutation, preserving repeated bootstrap draws, and evaluating layer choice separately.

\section{Conclusion}\label{sec:conclusion}

Linear probes read correctness-related information from all six diffusion code models tested. The confidence comparisons offer no consistent advantage for the probe, and the tested positive steering interventions provide no dependable gain. A next step is to target representations of particular errors and repairs, and to test when those representations become available during denoising.

\section*{Data availability}
The supplementary package includes the measurements used to draw the figures, experiment summaries, and per-task steering outcomes. The data guide maps these files to the reported results. Large activation arrays and model weights are not included.

\section*{Code availability}
The supplementary package contains the probe and steering scripts, figure-generation code, and instructions for using each component. The figures can be rebuilt from the included measurements without running a model. Repeating the model experiments requires additional inputs and dependencies, which are described in the reproduction guide.

\noindent Project repository: \href{https://github.com/efficientcomputation/heckler-in-the-hidden-state}{\nolinkurl{efficientcomputation/heckler-in-the-hidden-state}}.

\FloatBarrier

\clearpage
% Supplementary methods and results.
\appendix
\setcounter{figure}{0}
\setcounter{table}{0}
\renewcommand{\thefigure}{A\arabic{figure}}
\renewcommand{\thetable}{A\arabic{table}}
\renewcommand{\theHfigure}{appendix.\arabic{figure}}
\renewcommand{\theHtable}{appendix.\arabic{table}}

\section{Model comparison and vocabulary interpretation}\label{app:models}

Probe AUC measures discrimination between passing and failing programs. Vocabulary decoding identifies output tokens aligned with the probe direction.

The layerwise reconstructions peak at different layers from the original selections in five of six models. Table~\ref{tab:model_peaks} reports the original estimates separately from the curves in Figure~\ref{fig:depth}. Final-normalization conventions also differ; Appendix~\ref{app:final_layer} compares raw and normalized final-block states directly.

\begin{table}[htbp]
\caption{Peak estimates from the original six-model comparison.}
\label{tab:model_peaks}
\centering\small
\begin{tabular}{@{}llcc@{}}
\toprule
Model & Params & Original layer & Original AUC\\
\midrule
LLaDA-flash          & 102B         & L20 & 0.81\\
DiffuCoder-7B        & 7.6B         & L22 & 0.78\\
Dream-Coder-7B       & 7.6B         & L28 & 0.76\\
LLaDA2.0-mini        & 1.4B active  & L14 & 0.74\\
UltraLLaDA-8B        & 8B           & L27 & 0.72\\
Stable-DiffCoder-8B  & 8.25B        & L19 & 0.67\\
\bottomrule
\end{tabular}
\end{table}

\begin{figure}[htbp]
\centering
\includegraphics[width=.64\linewidth]{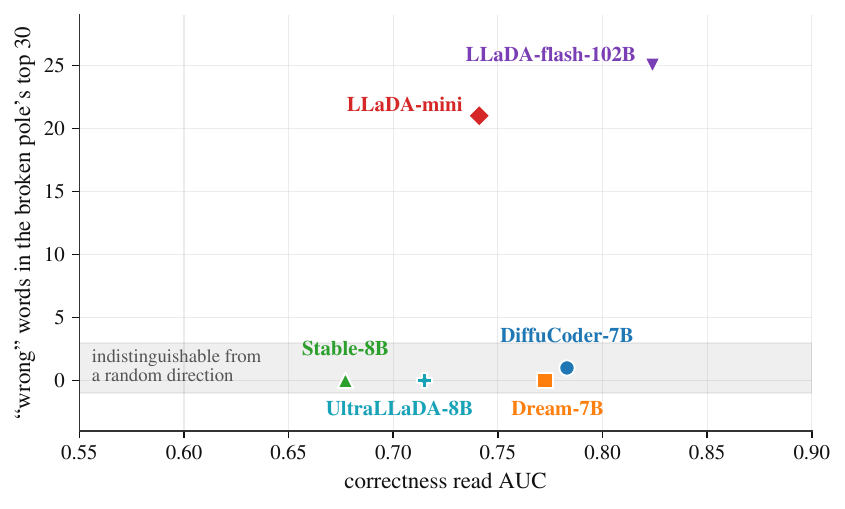}
\caption{Probe AUC (horizontal axis) and wrongness-word enrichment (vertical axis). All six models have above-chance readouts, but only the LLaDA2-MoE pair strongly enriches explicit ``wrong / incorrect'' words. The dense coders' vocabulary enrichment is near the random-direction baseline.}
\label{fig:interpretable}
\end{figure}

The LLaDA2-MoE models show stronger enrichment of wrongness terms than the dense coders despite comparable correctness readouts (Figure~\ref{fig:interpretable}).

\clearpage
\section{Capture protocol, scoring, and uncertainty}\label{app:methods}

Table~\ref{tab:cohorts} lists the analysis samples. Within-problem AUC uses tasks containing both passing and failing attempts. The family comparison and follow-up experiments use separate captures.

\begin{table}[htbp]
\centering\small
\caption{Problems and examples included in each analysis.}
\label{tab:cohorts}
\begin{tabular}{@{}p{.49\linewidth}rr@{}}
\toprule
Cohort & Problems & Examples\\
\midrule
DiffuCoder, 32 denoising steps & 19 & 95\\
DiffuCoder, 16 denoising steps & 54 & 324\\
Qwen2.5-Coder, generated attempts & 56 & 448\\
DiffuCoder, MBPP+ mutants & 52 & 392\\
DiffuCoder, HumanEval+ mutants & 44 & 334\\
LLaDA-flash / Stable remeasurement & 53 & 399 each\\
\bottomrule
\end{tabular}
\end{table}

Before mixed-problem filtering, the 32-step DiffuCoder sample contains 120 tasks with five attempts each, and the Qwen2.5-Coder sample contains 180 with eight attempts each. Analysis uses DiffuCoder's last-step response-token mean and Qwen's completion-token mean.

Activations are centered within problem. Five folds are assigned by problem using NumPy seed 0; attempts from the same problem never cross folds. The normalized difference between passing and failing training activations defines each fold's direction. AUCs of held-out projections are averaged equally across mixed problems. The implementation uses sorted ranks without explicit tie averaging. The maximum of the out-of-fold curve selects the reported layer, without a separate layer-selection holdout.

Diffusion confidence controls use mean visible-token log probability, negative mean visible-token entropy, and mean completion log probability with response positions masked. Diffusion logits at $i$ score token $i$; autoregressive logits score the following token. Visible and masked passes condition on different information.

\paragraph{Null distributions.}\label{app:uncertainty} We use 200 within-problem label permutations. For the confidence comparison, each permutation evaluates fixed out-of-fold scores and repeats layer selection; probe directions are not refitted. The perturbation analyses evaluate fixed scores at a preselected layer. Final-layer analyses select the peak before computing its null, so those peak scores do not account for layer selection.

\paragraph{Intervals.} The probe--confidence bootstrap implementation merges repeated problem identifiers after resampling, losing their sampling multiplicity. We therefore do not interpret its intervals as cluster-bootstrap confidence intervals. Correct intervals require resampling paired per-problem AUC differences, with layer selection addressed separately.

\clearpage
\section{Perturbation controls}\label{app:perturbations}

\begin{figure}[htbp]
\centering
\includegraphics[width=.60\linewidth]{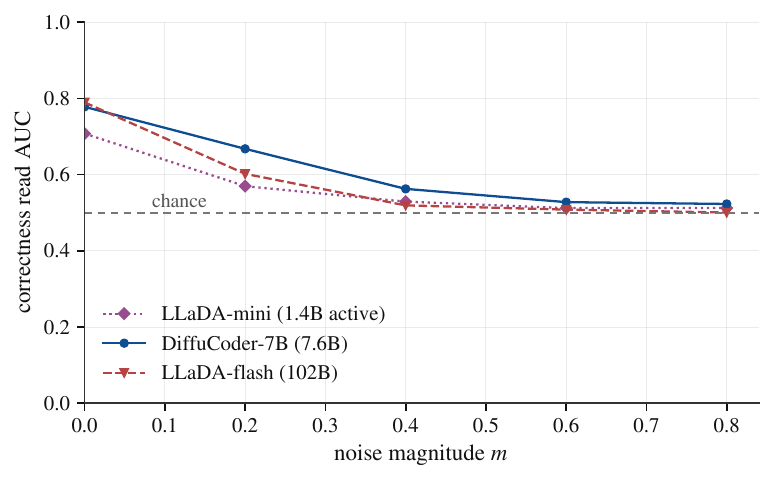}
\caption{Mean correctness-probe AUC across five noise realizations. Noise of relative magnitude $m$ is added at the probed layer. Performance declines at different rates across the three models.}
\label{fig:noise}
\end{figure}

DiffuCoder, LLaDA-mini, and LLaDA-flash use noise seeds $s=0,1,2,3,4$, mapped to NumPy seeds $1000000+s$. One random unit direction per example is shared across positions and scaled by the mean token-state norm. The remaining three models were evaluated with one noise realization. Figure~\ref{fig:noise} shows the three models with repeated seeds.

\begin{figure}[htbp]
\centering
\includegraphics[width=.62\linewidth]{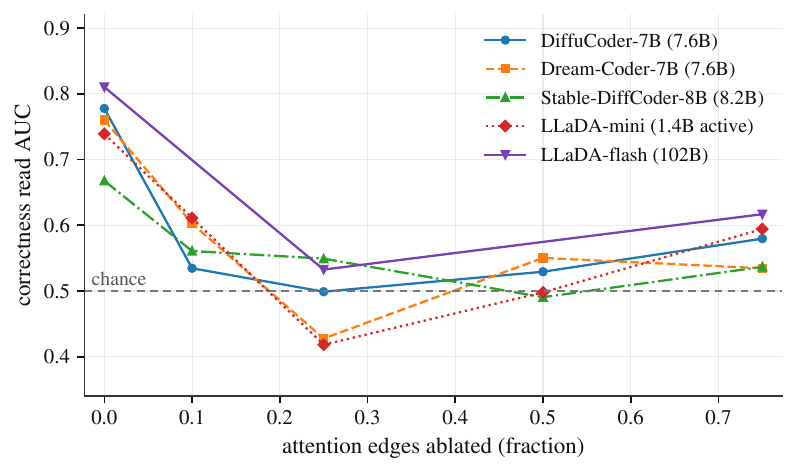}
\caption{Correctness-probe AUC after attention-edge removal. Most sub-chance dips and subsequent increases have per-point $|z|<2$ relative to their shuffled-label nulls. The curves use a single seed.}
\label{fig:causal}
\end{figure}

\clearpage
\section{Steering configuration and results}\label{app:steering}

The follow-up steering experiments intervene at DiffuCoder block index 18. For a unit direction $v$, we add $\alpha\bar r v$ to selected response positions at every denoising step, where $\bar r$ is the row's mean response-token residual norm and $\alpha=\pm0.4$. Global arms affect all response positions; conditional arms affect only positions still containing the mask token. All eleven arms share one generation batch: an untreated baseline, probe directions, random directions, and a learned direction.

A fresh random unit direction is sampled per task with NumPy seed $10000+i$, where $i$ is the evaluation-local task index. The evaluation configuration specifies 256 new tokens, 32 denoising steps, and temperature 0. Repeated untreated rows in the earlier bf16 sweep had different pass counts.

The learned direction starts from the probe and minimizes reference-token negative log probability from a masked-response input while model weights are frozen. Adam runs for 25 steps at learning rate 0.005, alternating between two groups of four references.

Table~\ref{tab:steerarms} reports results on 30 tasks. The learned direction has cosine similarity 0.108 with the probe direction. Its positive global intervention reduces passing programs from 15 to 5 and compiling programs from 26 to 9.

\begin{table}[htbp]
\centering\small
\caption{Passing and compiling programs out of 30 evaluation tasks.}
\label{tab:steerarms}
\begin{tabular}{@{}lrr@{}}
\toprule
Intervention & Pass tests & Compile\\
\midrule
Baseline & 15 & 26\\
Probe, global $+$ / $-$ & 14 / 5 & 24 / 16\\
Probe, conditional $+$ / $-$ & 15 / 10 & 25 / 21\\
Random, global $+$ / $-$ & 15 / 14 & 22 / 24\\
Random, conditional $+$ & 16 & 25\\
Learned, global $+$ / $-$ & 5 / 15 & 9 / 27\\
Learned, conditional $+$ & 7 & 12\\
\bottomrule
\end{tabular}
\end{table}

The main steering figure summarizes a separate sweep; its error bar is the standard deviation across ten random directions.

\clearpage
\section{Final-layer measurements and the compression hypothesis}\label{app:final}\label{app:final_layer}

\begin{figure}[htbp]
\centering
\resizebox{0.76\textwidth}{!}{%
\begin{tikzpicture}[font=\small]
  \draw[-{Stealth[length=2mm]}, soft, line width=0.9pt] (0,0) -- (11.4,0);
  \draw[-{Stealth[length=2mm]}, soft, line width=0.9pt] (0,0) -- (0,3.4);
  \node[soft, below] at (5.3,-0.1) {network depth (layer) $\rightarrow$};
  \node[soft, rotate=90, anchor=south, align=center] at (-0.45,1.7) {readability of the\\correctness signal};
  \draw[soft, dashed, line width=0.7pt] (0,0.45) -- (10.4,0.45);
  \node[soft, font=\scriptsize, above] at (1.1,0.45) {chance};
  \fill[emph!9] (8.7,0) rectangle (10.4,3.2);
  \node[emph, font=\scriptsize, align=center] at (9.55,-0.42) {final\\layer(s)};
  \draw[accent, line width=1.5pt] plot[smooth] coordinates
    {(0.2,0.47)(1,0.95)(2,1.3)(3.2,1.75)(4.5,2.2)(6,2.5)(7.2,2.55)(8,2.4)(8.8,1.75)(9.5,1.0)(10.15,0.62)};
  \fill[accent] (7.2,2.55) circle (2.2pt);
  \node[accent, font=\footnotesize, align=center] at (3.7,2.7) {built with depth\\(readable)};
  \node[emph, font=\footnotesize, align=center] at (9.55,2.75) {collapses?};
  \node[draw=ink!45, fill=ink!5, rounded corners=2pt, align=center, inner sep=5pt,
        text width=2.3cm, font=\footnotesize] (tok) at (12.7,1.5) {output token distribution};
  \draw[-{Stealth[length=2.4mm]}, soft, line width=1.0pt] (10.2,0.9) .. controls (11.2,1.2) .. (tok.west);
\end{tikzpicture}}
\caption{Schematic of the proposed compression mechanism, not measured data. A late conversion toward output-token directions might reduce a linear probe's access to correctness information. The DiffuCoder-7B tests below did not support this explanation.}
\label{fig:compression}
\end{figure}
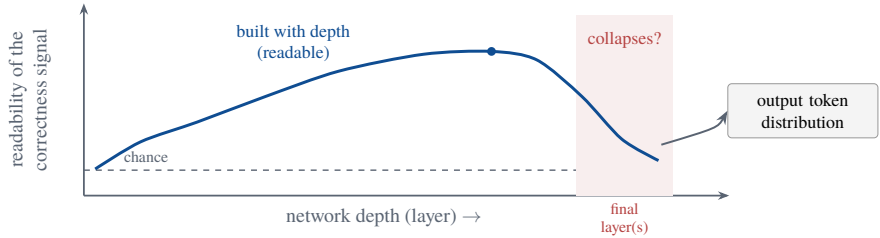

A transformer block's raw output and its final normalized state are different measurements. Their inconsistent treatment in the initial captures motivated separate measurements before and after final normalization. We also tested whether projection toward output-token directions could explain a late decline in probe AUC (Figure~\ref{fig:compression}).

The DiffuCoder final-layer battery uses five-fold splits by problem. Logistic regression uses $C=0.05$ and at most 2000 iterations. The multilayer perceptron has one 64-unit hidden layer, regularization parameter 0.01, at most 600 iterations, and random seed 0. The vocabulary-subspace test uses the leading 64 right singular vectors of the mean-centered output embedding matrix. The normalization test applies RMSNorm to an already pooled final-block vector; this is not generally equivalent to normalizing each token before pooling.

In that capture, the final-block linear AUC is 0.838 against a peak of 0.842; the normalized pooled-vector AUC is 0.816. Splitting the representation by the selected vocabulary subspace gives higher AUC outside it than inside it: 0.829 versus 0.781 at the peak, and 0.811 versus 0.741 at the final block. The tested nonlinear decoder does not yield a higher point estimate than the peak linear probe. DiffuCoder therefore retains a strong correctness read at its final block, without evidence for the proposed compression mechanism.

The LLaDA-flash and Stable remeasurements instead capture final-block token states directly, apply final normalization to each token, and then average. They use compilable reference-code mutants, with passing and failing mutants for each retained base problem. Inputs are truncated to 640 tokens and processed singly. LLaDA-flash has peak AUC 0.8002, raw-final AUC 0.6840, and normalized-final AUC 0.6976. For Stable-DiffCoder, the corresponding AUCs are 0.7301, 0.6086, and 0.6325.

Stable-DiffCoder also shows a final-layer decline. The effect is therefore not confined to the 102B model, and these comparisons do not establish a scale threshold.

\clearpage
\section{Diffusion versus autoregressive models}\label{app:comparison}\label{app:architecture}

\begin{figure}[htbp]
\centering
\includegraphics[width=.64\linewidth]{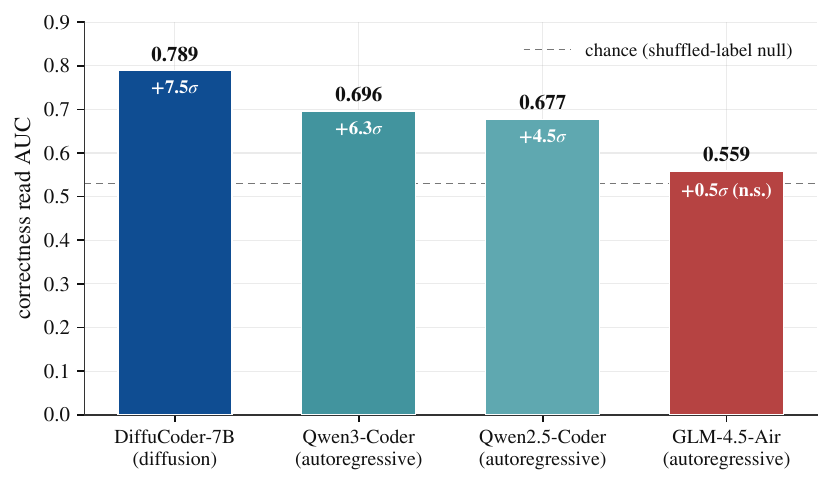}
\caption{Correctness probes across diffusion and autoregressive models. Each model is scored against its own shuffled-label null; the displayed $\sigma$ values describe distances from those separate nulls and should not be compared across bars as a common effect size. Task sources and training also differ.}
\label{fig:comparison}
\end{figure}

Is a readable correctness signal special to diffusion language models? We applied the same type of linear probe to two autoregressive code models on SWE-bench-derived rollouts. One of them, Qwen3-Coder, reads nearly as well as the diffusion models in this comparison, with an AUC of 0.70. The other, GLM-4.5-Air, is much closer to its shuffled-label baseline, with an AUC of 0.56 (Figure~\ref{fig:comparison}).

We later added Qwen2.5-Coder on MBPP+, the programming benchmark used in the diffusion experiments. Its probe reaches an AUC of 0.677, while its own log-probability confidence reaches 0.703. The probe peaks near the end of the layer stack, whereas DiffuCoder's peak appears earlier. Their training histories, task samples, and capture settings differ, so the comparison does not isolate an effect of architecture.

During diffusion denoising, a correctness signal can be present while multiple positions in the program remain revisable. In an autoregressive model, the tokens already written are committed, and a read taken mid-generation can only shape what comes next. Diffusion therefore offers an opportunity to use information about the program while the answer can still change. Whether that opportunity is useful depends on when the signal appears and whether an intervention supplies a repair rather than just a correctness verdict.

In the matched confidence comparisons, the probe offers no consistent advantage over the model's own confidence. Nor did positive probe or learned-direction steering improve pass counts in the 30-task evaluation. The tested models provide a readable signal; the interventions leave open how to turn that signal into a useful repair.

\end{document}